# Irradiation induced vacancy creation in single walled carbon nanotubes

Sumera Javeed[1], Sumaira Zeeshan[1], Shoaib Ahmad[1,2*]

[1]PINSTECH, P.O. Nilore, Islamabad, Pakistan

[2]National Centre for Physics, Quaid-i-Azam University Campus, Shahdara Valley, Islamabad, 44000, Pakistan

[*]Email: sahmad.ncp@gmail.com

## Abstract

Single walled carbon nanotubes of 2nm diameter and 3-13 μm length were compressed in Cu bullets and irradiated with $Cs^+$ ions with energies from 0.2 to 2.0 keV and subjected to successively increasing $Cs^+$ dose. From the mass spectra of the sputtered carbon atoms and clusters as a function of $Cs^+$ energy the monitoring of trend of the relative number densities of the fragmenting species indicates the accumulating damage. Irradiation induced fragmentation provide clues to the structural changes as a result of creation of vacancies due to the sputtering of mono, di-, tri-atomic and higher species. Monitoring of the sample's electrical conductivity provides in situ information of the sequences of the transformation that may be occurring in the structures of the nanotube as a function of $Cs^+$ energy and dose.

## 1. Introduction

Electron and ion irradiations of single walled carbon nanotubes (SWCNTs) lead to understanding of the defect production mechanisms and defect-related property changes [1-3]. The relative ratios of the defects produced like single versus di-vacancy and the role that displaced and knocked-off atoms, diatoms may play in the formation of new structures within and between the nanotubes has been simulated and experimentally studied [4-6]. Our recent work on the irradiated fullerite has identified the structural transitions that occur due to fragmentation of the $C_{60}$ cages as a function of $Cs^+$ energy $E(Cs^+)$ [7]. Prolonged $Cs^+$ irradiation led to cages' destruction. Similarly, sputtering profile of the clusters from irradiated multi walled carbon nanotubes (MWCNTs) was studies as a function of the ion dose that leads to structural transformation [8]. Energy of the irradiating ion determines the nature and extent of the damage to carbon's nano structures. High doses of heavy energetic ions may induce structural changes in nanotubes that vary from fusion and welding of nanotubes to amorphization. Large scale sputtering from a collection of compressed single walled carbon nanotubes may have twin effects;

(1) single and di-vacancies in the nanotubes might open up the cylinders of tubes in localized regions and

(2) the consequent re-deposition of these clusters on the neighbouring nanotubes could initiate new nano structures in and around the nanotubes.

We have observed transitions in the irradiated fullerite and MWCNTs [7,8]. In this paper we report results from an extended study on SWCNTs; the similarities and differences are pointed out.

## 2. Materials and Methods

This investigation is primarily based on mass spectrometry of the sputtered atoms and clusters $C_x$ (x=1- 6), under increasing energy and dose of $Cs^+$ irradiations on SWCNTs of 2 nm diameter and 3-13 μm length that were compressed into 4 mm diameter x 2 mm deep hole in Cu bullets. The bullets containing the SWCNTs are used as the target in the source of negative ions with cesium sputtering (SNICS). $Cs^+$ energy with respect to the target containing SWCNTs was varied from 0.2-2.0 keV in 0.1 keV steps. In SNICS one can vary $Cs^+$ energy while keeping the sputtered species' energy constant by adjusting the extraction voltage. Sputtered $C_x$ were accelerated as anions at constant beam energy of 30 keV. The dose effects are investigated by comparing the variations of the sputtered species' relative yields and the cumulative effects of $Cs^+$ dose. SNICS is employed as the experimental tool to induce SWCNT fragmentation as a result of $Cs^+$ bombardments and monitored the sputtered $C_x$. Higher electron affinities of $C_x$ facilitate anion formation and detection. A 30° magnet was used for the mass analysis of anions.

# 3. Results

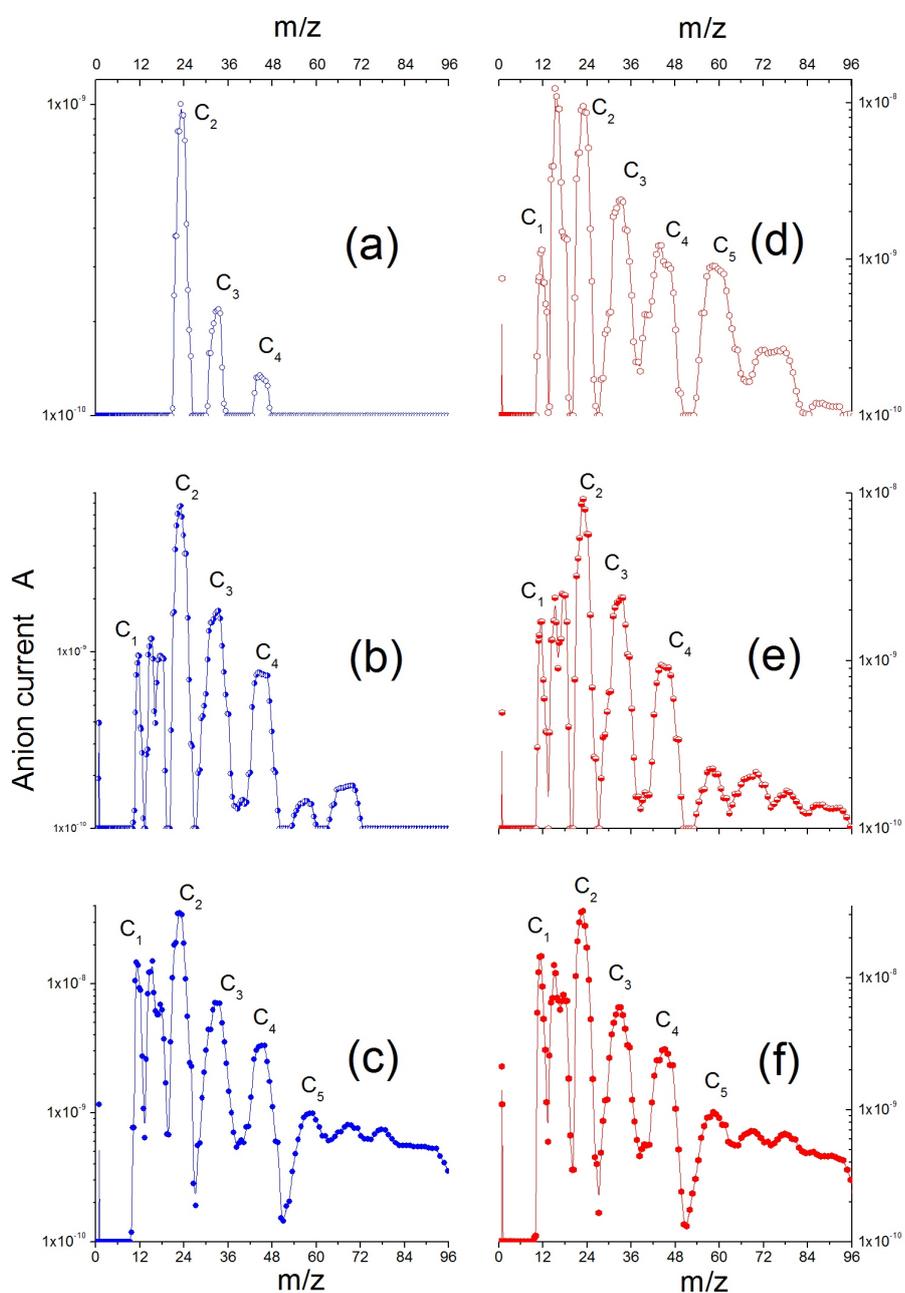

Fig. 1. Two sets of mass spectra of $C_x$ sputtered from the irradiated SWCNTs are shown. The first set Fig. 1(a) to (c) shows $C_x$ emitted from the pristine sample while Fig1(d) to (f) are from the Cs rich, restructured SWCNTs at $E(Cs^+)$=0.2, 0.6 and 2.0 keV respectively.

## 3.1. The $C_2$ dominated mass spectra

Two sets of m/z spectra of $C_x$ anions were obtained as a function of $E(Cs^+)$; one from the pristine and the other were repeats in the same energy range on the Cs-rich sample of SWCNTs. Nineteen m/z spectra of $C_x$ from each set were obtained. Fig. 1 shows three selected spectra from each set where 3(a)-(c) are for the pristine and (d)-(f) for the Cs rich SWCNTs. Fig. 1(a) is the first mass spectrum from the pristine SWCNTs showing only $C_2$, $C_3$ and $C_4$. These three species retain the bulk of $C_x$ output for $Cs^+$ irradiations at all energies with relative variations in intensities. $C_2$ is present in all the spectra as the most prominent fragment. $C_1$ makes its first appearance at $E(Cs^+)$ = 0.4 keV and steadily rise to about 10%. In the Cs-rich SWCNTs shown in Fig. 1(d) at $E(Cs^+)$ = 0.2 keV, $C_1$, $C_5$ and impurities like O and $H_2O$ are also sputtered in addition to $C_2$, $C_3$ and $C_4$. This is in sharp contrast with Fig. 1(a) at the same $E(Cs^+)$. Similar is the case for Fig. 1(e) and (f).

## 3.2. Normalized yields of the sputtered $C_x$

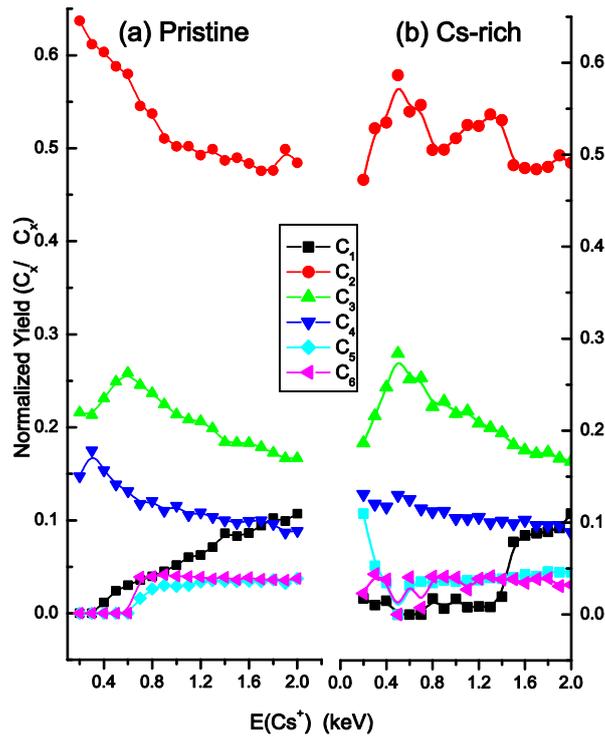

Fig. 2. The normalized yield $C_x/\Sigma C_x$ of $C_1$, $C_2$, $C_3$, $C_4$, $C_5$ and $C_6$ are plotted as a function of $E(Cs^+)$ from 0.2 to 2.0keV; from the pristine sample in 2(a) and from the Cs-rich sample in 2(b).

Fig. 2 has the normalized yields $C_x/\Sigma C_x$ for x=1 to 6 of the sputtered $C_x$ from the pristine SWCNTs and after 100 minutes of $Cs^+$ bombardment. The normalized plots provide us the landscape of the comparative densities. $C_2$, $C_3$ and $C_4$ are the main species emitted from the irradiated SWCNTs with subtle differences between their number densities from the pristine set of nanotubes and those from the heavily irradiated ones. These differences illustrate the nature of the damage to SWCNTs and the consequent emergence of the sputtered $C_x$-initiated structures. Fig. 2(a) shows $C_2$'s relative yield decreasing from 64% at $E(Cs^+)=0.2$ keV to ~50% between 1 and 2 keV. $C_3$'s yield increases from 20% to a broad peak of ~25% at 0.6 keV and stabilizes around its initial yield. $C_4$ is a stable species with ~10% relative yield for the entire $E(Cs^+)$ range. The relative population of $C_1$ increases from 0 to ~10% indicating a uniform increase in the production of single vacancies with $E(Cs^+)$. The large clusters $C_5$ and $C_6$ are not detectable at $E(Cs^+) < 0.5$ keV but once formed, these retain their steady share of ~3-5%. Fig. 2(b) shows similar broad feature of cluster emission profile that is obtained after 100 minutes' irradiation of the SWCNTs. Here again, $C_2$, $C_3$ and $C_4$ are the main constituents emitted from the $Cs^+$ sputtered nanotubes. $C_3$'s yield is similar to that in Fig. 2(a). $C_4$ also has a steady 10%. $C_1$ is not sputtered from the heavily irradiated ensemble of SWCNTs until ~$E(Cs^+)=1.5$ keV, after which its share is steady around 10% indicating $C_1$ as a by-product of the larger $C_x$ fragmentation.

## 3.3. Enhancement of the electrical conductivity of the irradiated SWCNTs

Target current variations with $E(Cs^+)$ are shown in Fig. 3. The pristine sample has a steady conductivity $\sigma \sim 2.5 \times 10^{-8}$ $AV^{-1}$ from 0.2 to 2.0 keV. Cs rich SWCNTs have two regimes; one for 0.1-1.4 keV with $\sigma \sim 2.1 \times 10^{-8}$ $AV^{-1}$ the other has a sharp increase for 1.4-2.0 keV with $\sigma \sim 2 \times 10^{-7}$ $AV^{-1}$. The measured conductivity of SWCNTs that are compressed as bundles of ropes occurs through their surface interactions. Low $\sigma$ may indicate the state of the undamaged nanotubes. Even with Cs rich SWCNTs, the structural transformation seems to occur after $E(Cs+) \geq 1.4$ keV. The presence of the inter-SWCNT structures may have been initiated after the partial opening up of the nanotubes after removal of extensive amount of carbon in the form of atoms and clusters.

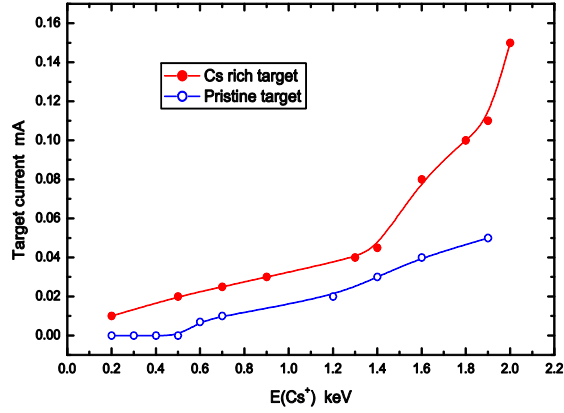

Fig. 3. The increase in the target current as a function of the $E(Cs^+)$ and dose is shown. The initial sequence on the pristine sample is shown with open and the one on the Cs-rich is shown with filled circles.

## 4. Discussion

Energy dissipating collisions of $Cs^+$ with SWCNTs and their constituents has certain similarities with those in MWCNTs and fullerite. The mechanisms of the generation of primary recoils are similar in fullerite, and nanotubes, the relative yields of the collision products may be different in each case. In the case of the irradiated MWCNTs $C_1$ and $C_2$ are the major byproducts of direct $Cs^+$-$C_1$ and $Cs^+$-$C_2$ collisions [9]. $Cs^+$-$C_{60}$ collisions in fullerite occur in different energy regimes; below 400 eV only $C_2$s are emitted while for higher $E(Cs^+)$ all species are emitted [8]. This effect is unique to the irradiated fullerite and has not been observed in graphite, single or multi walled carbon nanotubes. From the irradiated SWCNTs $C_2$ is the most intense species with $C_1$ being virtually absent; $C_3$ is the next competing in intensity cluster with $C_4$ emitted as a stable cluster. Collision cascades are likely to occur in MWCNTs due to the availability of nearest neighbours of a primary knock-on atom. This does not seem likely to occur in SWCNTs. The relative absence of $C_1$s and the associated single vacancies as opposed to the predominance of $C_2$s with the generation of di-vacancies are unique to the irradiated SWCNTs. The higher energy required for the creation of single vacancies $E_{sv}$ ~5-7eV as opposed to $E_{dv}$ ~3-5 eV for a di-vacancy may be the primary reason [11]. With $E_{sv}>E_{dv}$ the probability of finding $C_1$s as compared with $C_2$s among the sputtered species ~ $\exp(-E_{sv}/E_{dv})$; implying the number densities $N_{sv}/N_{dv}$ ~ 10-15%. This is the approximate ratio of the sputtered $C_1$:$C_2$ in our experiments. $C_2$ with the highest relative yield at all $Cs^+$ energies together with $C_3$ and $C_4$ is present even at the smallest $E(Cs^+)$~0.2 keV.

# 5. Conclusions

$C_2$, $C_3$ and $C_4$ are the most persistent sputtered species from the irradiated SWCNTs. $C_1$ is the least intense of all the species. We interpret it as the ease with which di-vacancies can be created as opposed to the single vacancies. We also propose that tri- and double di-vacancies may have been produced in the irradiated SWCNTs. The large number densities of the sputtered clusters $C_x$ may be involved in the formation of networks of chains, rings and sheets that interconnect different SWCNTs. The accumulation of the sputtered species in the inter-SWCNT space may be responsible for the increased connectivity among the nanotubes and result in the consequent increase in the electrical conductivity that was observed. Fragmentation of the SWNTs may lead to the destruction of the nanotube structure locally leading to the opening of the tubes. The nature and extent of the cumulative irradiation induced damage is to fragment the individual nanotube structure while simultaneously bonding or welding these with each other. The energetic ion irradiation effects are destructive for the nanotube structure and at the same time a reconstructive process occur due to the accumulation of the sputtered clusters. The irradiation induced structural transitions is seen to occur with these two simultaneously operative processes.